\documentclass{WileyMSP-template}

\usepackage{titlesec}%
\usepackage{subfigure}%
\usepackage{amsmath}%
\usepackage{amsfonts}%
\usepackage{amssymb}%
\usepackage{feynmp}%
\usepackage{graphicx}%
\usepackage{setspace}%
\usepackage{comment}%
\usepackage{dsfont}%
\usepackage{color}%
\usepackage{bbold}
\usepackage{tikz}
\usetikzlibrary{arrows,calc}
\usepackage[pdftex,colorlinks=true,linkcolor=blue,citecolor=blue]{hyperref}
\usepackage{ulem}

\graphicspath{{fig/}}

\newcommand{\imag}{\mathrm{i}}
\newcommand{\dif}{\mathrm{d}}
\newcommand{\lambdabar}{{\mkern0.75mu\mathchar '26\mkern -9.75mu\lambda}} 

\newcommand{\rbrac}[1]{\left[ #1\right]}
\newcommand{\of}[1]{\left( #1\right)}
\newcommand{\abs}[1]{\left| #1\right|}

\newcommand{\dwg}{d_\mathrm{wg}}
\renewcommand{\vec}[1]{\textbf{\textit{#1}}}
\usepackage{import}
\usepackage{graphics}
\usepackage{siunitx}
\usepackage[super]{nth}
\begin{document}

\pagestyle{fancy}
\rhead{\includegraphics[width=2.5cm]{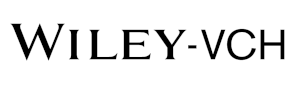}}

\title{Broadband mode division multiplexing of OAM-modes by a micro printed waveguide structure}

\maketitle


\author{Julian Schulz}
\author{Georg von Freymann}

\begin{affiliations}
Julian Schulz\\
Address Physics Department and Research Center OPTIMAS, RPTU Kaiserslautern Landau, 67663 Kaiserslautern, Germany\\
Email Address: schulzj@rhrk.uni-kl.de\\
ORCID: 0000-0003-4630-4117\\
Prof. Dr. Georg von Freymann\\
Address Physics Department and Research Center OPTIMAS, RPTU Kaiserslautern Landau, 67663 Kaiserslautern, Germany\\
Adress Fraunhofer Institute for Industrial Mathematics ITWM, 67663 Kaiserslautern, Germany\\
Email Address: georg.freymann@rptu.de\\
ORCID: 0000-0003-2389-5532\\
\end{affiliations}



\begin{abstract}
A light beam carrying orbital angular momentum (OAM) is characterized by a helical phase-front that winds around the center of the beam. 
These beams have unique properties that have found numerous applications. In the field of data transmission, they represent a degree of freedom that could potentially increase capacity by a factor of several distinct OAM modes. While an efficient method for (de)composing beams based on their OAM exists for free-space optics, a device capable of performing this (de)composition in an integrated, compact fiber application without the use of external active optical elements and for multiple OAM modes simultaneously has not been reported.
In this study, a waveguide structure is presented that can serve as a broadband OAM (de)multiplexer. The structure design is based on the adiabatic principle used in photonic lanterns for highly efficient conversion of spatially separated single modes into eigenmodes of a few-mode fiber. In addition, an artificial magnetic field is introduced by twisting the structure during the adiabatic evolution, which removes the degeneracy between modes having the same absolute OAM. This structure can simplify, stabilize, and miniaturize the creation or decomposition of OAM beams, making them useful for various applications.
\end{abstract}


\section{Introduction}
The special properties of light beams with orbital angular momentum (OAM) have enabled significant advances in astrophysics \cite{Astronomical_vortex_coronagraph}, high resolution microscopy \cite{Overcoming_Rayleigh_Limit}, remote sensing \cite{Detection_of_Spinning_Object}, optical tweezers \cite{twisted_tweezers} and many more. 
In particular, the field of mode division multiplexing has sparked interest in the mode space of OAM modes to address the exponentially increasing demand for data transmission capacity \cite{SDM_in_optical_fibres,SDM_Transmission_Systems}. 
A light beam carrying orbital angular momentum $\ell\in\mathds{Z}$ has a cross section where the orbit around the beam center acquires a phase of $2\pi\ell$. 
For $\ell\neq0$, the phase singularity in the beam center causes the intensity to drop to zero.
OAM beams with different $\ell$ are orthogonal and thus allow the OAM to be used as an identifier of different channels to transport an increased amount of information through a single fiber. 
Other independent properties of light, such as wavelength and polarization, have been used to increase transmission capacity by multiplexing.

Multiple optical components can generate OAM beams. 
The phase of an expanded beam can be altered using spiral phase plates, spatial light modulators, or diffractive phase holograms. 
A highly effective technique for efficiently generating and decomposing OAM beams was introduced in Ref.\cite{Efficient_OAM-Sorting}, which uses a log-polar transformation performed by two fixed optical elements.
Metasurfaces and gratings can be designed to carry OAM in the scattered light, but with a limited conversion ratio to the input beam. 
These components are designed specifically for the wavelength used to imprint the desired phase shift. 
In addition to techniques based on a fixed spatial phase relationship, q-plates can generate OAM beams from defined spin angular momentum (SAM) beams based on a medium with strong OAM-SAM coupling.

These spatial methods are well-established for generating OAM beams in free space. However, for transmitting data in fibers, they are not without drawbacks. 
Reductions in device volume can lead to decreased mode purity, making miniaturization challenging.
Due to the high refractive index difference between air and fiber, fiber coupling results in losses\cite{Review_OAMGeneration_in_Fibers}. Additionally, even slight lateral misalignment can lead to crosstalk between modes in fiber\cite{OAM_of_light_for_communications}.
To solve these issues in fiber communication, techniques for generating and decomposing OAM within waveguides and fibers have been proposed and implemented.

Twisted spiral waveguides\cite{chiral_fiber_gratings,coupled_elliptical_fiber} or helical waveguides\cite{layered_helical_waveguides,long-period_helical_fiber} can be used to transform a the ground mode into an OAM mode. 
If the twist of these structures is chosen at the right frequency, the ground mode and one OAM mode have the same propagation constant and can couple effectively.
Coupled waveguide structures can be designed to clone, invert\cite{OAMsupermodes} or couple different\cite{Vector_Vortex_Beam_Emitter} OAM modes.
By lifting the degeneracy of the LP-modes for a fixed propagation length, those modes can be converted to OAM-modes by acquiring a phase-shift of $\pi/2$\cite{trench_silicon_waveguides,generation_via_vortex_fiber,Mode-Selective_PL,Modeselective_PL_and_polarization_controller}.
Multiple coherent input beams, controlled by external active optical elements for phase and amplitude, can also create OAM modes\cite{multiple_coherent_inputs,OAM_generation_based_on_PL,PL_OAM_Mux}.
In general, waveguide structures can achieve high efficiency, high mode purity, and a wide bandwidth, but only for one or a few OAM modes at a time or in a structure and only for low values of $\ell$ \cite{twisted_light_from_on-chip_devices}. 

\section{Principle of the structure}
In this study, we demonstrate a static waveguide structure capable of (de)multiplexing numerous OAM-modes simultaneously by arranging a ring core waveguide into five single mode waveguides, all without the use of active optical elements. 
Thus, the structure can be used for the superposition not just of different OAM modes but also at different wavelengths simultaneously.
Our findings indicate the effectiveness of this waveguide structure in achieving (de)multiplexing of OAM-modes and suggest possible avenues for future research. 
As a proof of principle, our structure successfully multiplexed OAM-modes with an absolute value of $\ell \leq 2$, but it has the potential to scale up and multiplex even higher modes. 

\begin{figure}
\def\svgwidth{\columnwidth}
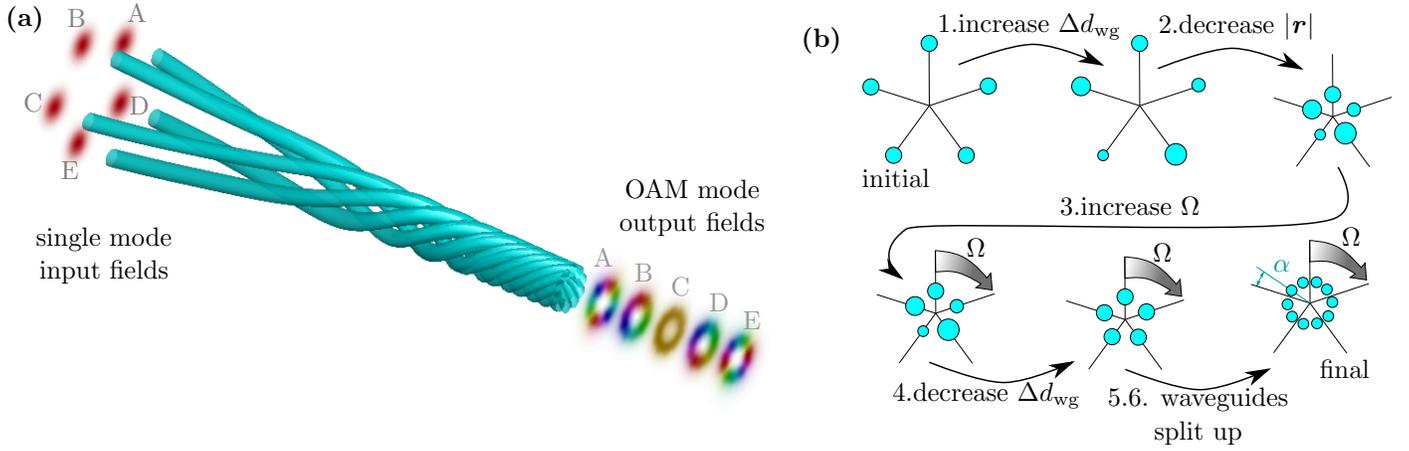
\caption{(a) 3D-Model of the waveguide structure. Dependent in which single-mode waveguide at the input facet light is coupled in, the light will be transformed in a OAM-state with a different $\ell$ as it reach the output facet. (b) schematic steps of the structure during the adiabatic evolution.}
\label{fig:modell_sketch}
\end{figure}

The main principle of this structure is the adiabatic transformation of eigenmodes from spatially separated single modes into modes in a ring waveguide carrying OAM. 
In an adiabatic evolution, the population of the eigenmodes remains constant while the eigenmodes change according to the system. 
Therefore, the change of the system should be significantly slower than the dynamics of the eigenmodes, as determined by the difference in propagation constants between the modes. 
Two mechanisms are utilized to maintain the propagation constants of each individual mode consistently spaced - individual waveguide detuning and an artificial magnetic field. 
Figure~\ref{fig:modell_sketch} depicts the resulting structure and the progression from localized eigenmodes to OAM-modes.

The propagation constant of a single waveguide can be detuned individually either by changing the refractive index of the core or, as in our case, by changing the diameter of the waveguide. 
When small changes are made, the propagation constant shows an almost linear dependence on the waveguide's diameter. 
The distinct propagation constants from differently sized fiber cores in photonic lanterns have already shown this selectivity feature with a high mode purity\cite{Six_mode_spatial_multiplexer,Mode-Selective_PL,Modeselective_PL_and_polarization_controller,OAM_generation_based_on_PL,PL_OAM_Mux}.
The paraxial Helmholtz equation can describe the evolution of the scalar transverse field amplitude $\psi$ along the propagation direction $z$ in these waveguide structures,
\begin{equation}
\label{eq:paaxialHelmholz}
\imag\lambdabar\partial_z\psi
=\frac{\lambdabar^2}{n_c}\nabla_\bot^2\psi-\Delta n\of{\vec{r}}\psi
\end{equation}
where $\lambdabar$ is the wavelength divided by $2\pi$, $n_c$ is the refractive index of the cladding material and $\Delta n$ is the local change of the refractive index.

Although the eigenvalues and eigenmodes of Equation~\ref{eq:paaxialHelmholz} can be adjusted by modifying the diameters of the waveguides individually, the eigenmodes of straight waveguide structures are always real-valued functions because of the system's time-reversal symmetry. 
To obtain eigenmodes that can only be expressed by complex-valued functions, a constant artificial magnetic field is introduced. This field distinguishes between modes with $+\ell$ and $-\ell$ and lifts their degeneracy.
To achieve this, the system must rotate as the light travels through the structure\cite{twist-induced_Aharonov-Bohm,Excitation_of_OAM_in_PCFibers}. 
This is described by the coordinate transformation: 
\begin{align}
\label{eq:basistransformation}
x'
&=x \cos\of{\Omega z} - y \sin\of{\Omega z} \\
y'
&=x \sin\of{\Omega z} + y \cos\of{\Omega z} \\
z'
&=z
\end{align}

Here $\Omega$ is the angular velocity or in a geometric interpretation, $2\pi/\Omega$ is the pitch of the helical trajectory of a waveguide.
The effects on the paraxial Helmholtz equation due to this coordinate transformation in the rotating frame of reference can be summarized in a constant vector potential $\vec{A}=n_c \Omega/\lambdabar \rbrac{-y,x,0}^T$ and an additional harmonic potential\cite{coreless_PCFiber}. 

\begin{align}
\label{eq:ArteficialMagneticField}
\imag\lambdabar\partial_z\psi
&=\frac{\lambdabar}{2n_c}\rbrac{\imag\nabla_\bot-\vec{A}}^2\psi
-\Delta n\of{\vec{r}}\psi
-\frac{n_c\Omega^2}{2}\rbrac{\hat{x}^2+\hat{y}^2}\psi \\
&=\frac{\lambdabar^2}{2n_c}\nabla_\bot^2\psi
+\frac{\lambdabar}{2n_c}\vec{B}\cdot\hat{L}\psi
-\Delta n\of{\vec{r}}\psi
\end{align}

It can also be expressed as the dot product of a constant magnetic field $\vec{B}=2n_c\Omega\lambdabar \vec{e}_z$ and the angular momentum operator $\hat{L}=-\imag\lambdabar\nabla_\bot\times\hat{r}$ acting on the state.
The synthetic magnetic field links the orbital angular momentum of the state with its propagation constant, and thus resolves the degeneracy of modes with equal absolute OAM. 
The distance between modes correlates with both the angular velocity $\Omega$ and the $\ell$ of the mode\cite{Helically_twisted_PCFibers}. 
It is important to note that the angular velocity cannot be increased without limit as the bound modes will scatter out of the structure due to the harmonic potential, which increases proportionally with $\Omega^2$. 

The resulting structure and the steps starting from localized eigenmodes and ending at OAM-modes are sketched in Figure~\ref{fig:modell_sketch}.

1.	The difference in diameter $\Delta \dwg$ of the waveguides is increased. The separated single-mode-waveguides are detuned against each other, the eigenmodes are localized, and the eigenenergies are gapped.

2.	The distance of the waveguides to the center is decreased. The light can couple between the waveguides and the eigenmodes spread over multiple waveguides.

3.	Angular velocity is increased. The degeneracy of modes with equal absolute $\ell$ is lifted. 

4.	The difference in diameter $\Delta \dwg$ of the waveguides is decreased again. For all eigenmodes, the intensity in the waveguides is evenly distributed, because the waveguides are no longer detuned. The energy gaps are only caused by the non-zero angular velocity.

5.	The waveguides split up to approach a ring waveguide.

6.	The diameter of the waveguides $\dwg$ is decreased. The thickness of the ring waveguide is reduced to filter out unwanted higher OAM-Modes \cite{multiple_coherent_inputs}. 

The perfectly round ring waveguide should be avoided due to its invariant structure under rotation, which restores the degeneracy of $+\ell$ and $-\ell$ OAM-modes and can result in crosstalk. 
Additionally, the evolution process does \underline{not} require the completion of one step before starting the next, allowing for partial overlap in time and reducing the total device length. 
To execute an adiabatic step, we employ a function resembling the Fermi-Dirac statistics for a parameter $s$ to gradually transition from an initial value $s_\text{i}$ to a final value $s_\text{f}$ over the propagation distance, $L$. 
The coefficients $\mu_s$ and $T_s$ are utilized to arrange the steps chronologically and establish their relative speed, correspondingly.

\begin{equation}
\label{eq:FermiDirac}
s\of{z}:\rbrac{0,L}\rightarrow\mathbb{R},
z\mapsto s_\text{i}+\frac{s_\text{f}-s_\text{i}}{\exp\of{\frac{z/L-\mu_s}{T_s}}+1}
\end{equation}

\section{Numerical calculation}
To demonstrate the general function and capabilities of the structure, such as mode purity and wavelength independence, scalar split-step BPM simulations were performed. 

The relative intensity $I_{\ell',\ell}$ of the mode with OAM $\ell'$ in the field at the output facet $\psi_{\ell}$  from the simulation data is calculated by an overlap integral\cite{multiple_coherent_inputs,trench_silicon_waveguides,OAM_generation_based_on_PL}. 
The index $\ell$ here represents the OAM into which the light from this input waveguide is supposed to be manly converted to.
\begin{equation}
I_{\ell',\ell}
=\rbrac{\frac{1}{\sqrt{\iint\psi_{\ell}\cdot\psi^*_{\ell} \dif x\dif y}}
\iint\psi_{\ell}\cdot\exp\of{-\imag \ell' \arg\of{x+\imag y}}\dif x \dif y}^2
\end{equation}
The functions $\exp\of{-\imag \ell' \arg\of{x+\imag y}}$ with $\ell'\in\mathds{Z}$ form an orthogonal basis that only acts on the phase winding of the field and ignores the radial distribution. 
Since the modes have only trivial radial structure this basis suffices for determining the relative intensity of modes and with that the mode purity $I_{\ell,\ell}$ in our simulations. 
The mode crosstalk $\sigma_\ell$, the ratio of intensity which is not converted into the desired mode with OAM $\ell$ to the total intensity, is then given by $\sigma_\ell=1-I_{\ell,\ell}$.

\begin{figure}
\def\svgwidth{\columnwidth}
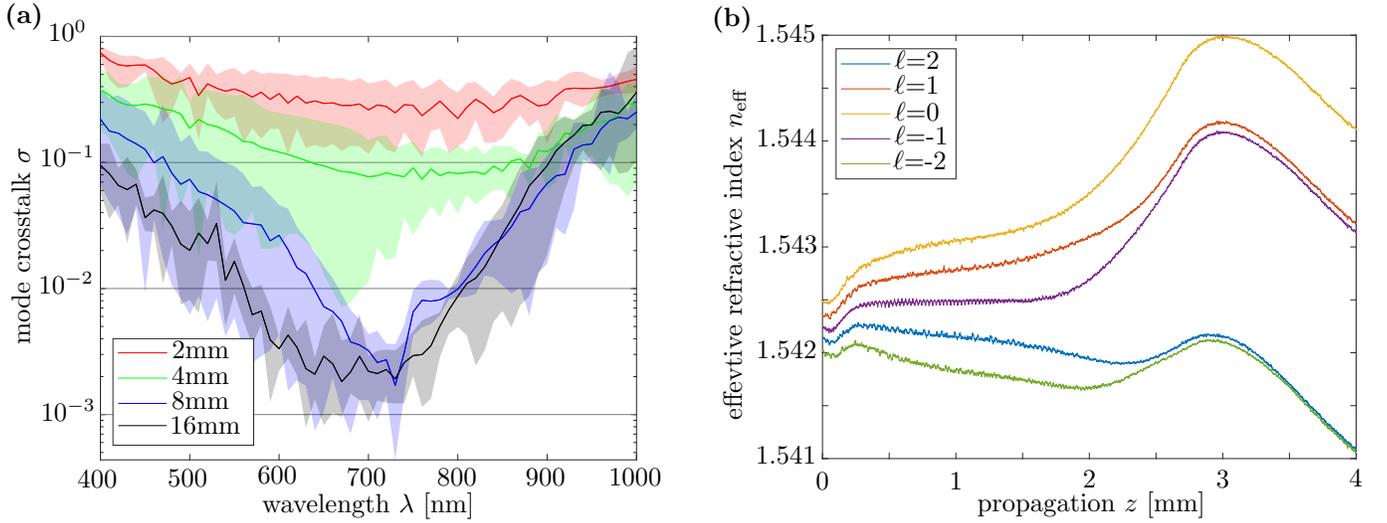
\caption{Simulation results. (a) Efficiency of the structure for different lengths. The mode crosstalk $\sigma$ is the ratio of intensity which is not converted into the desired mode to the total intensity. The line is the average crosstalk of the five modes while the borders of the transparent areas show the best and the worst mode conversion. With structure length increases adiabaticity and with it the mode purity. (b) Effective refractive index for each mode along the propagation.}
\label{fig:simulation}
\end{figure}

In general, for adiabatic evolution, numerous parameters such as coupling strength, potential depth, and magnetic field strength do not need to be set precisely. 
If the system evolves slowly and the propagation constants are sufficiently gapped, the final state changes only slightly. 
This inherent tolerance allows our device to operate effectively across a wide range of wavelengths. 
According to the adiabatic principle, increasing the length of the device slows down evolution and results in improved performance and a wider range of usable wavelengths. This can be observed in Figure~\ref{fig:simulation}a. 

Besides these quantities the effective refractive index $n_\mathrm{eff}$ over the propagation length for each eigenmode can be extracted form the simulations.
Since the phase acquired over one simulation step $\Delta z$ is proportional to the effective refractive index, the phase of the overlap integral of the field of two following steps is proportional to the effective refractive index.
\begin{equation}
\label{eq:EffectiveRefractiveIndex}
n_\mathrm{eff}
=\frac{\lambdabar}{\Delta z} \arg\of{\iint\psi\of{z}\cdot\psi^*\of{z+\Delta z} \dif x \dif y}
\end{equation}

As shown in Figure~\ref{fig:simulation}b the effective refractive index is kept gapped most of the time till the end where a more rotational symmetric structure is approached.

\section{Comparison with the experiment}

To experimentally demonstrate the device, we fabricated corresponding waveguide structures, following principle used in\cite{jorg_artificial_2020,NNNSchulz}. 
The sample is fabricated out of IP-Dip (Nanoscribe GmbH) using the commercial direct laser writing (DLW) system Photonic Professional GT from Nanoscribe.

\begin{figure}
\def\svgwidth{\columnwidth}
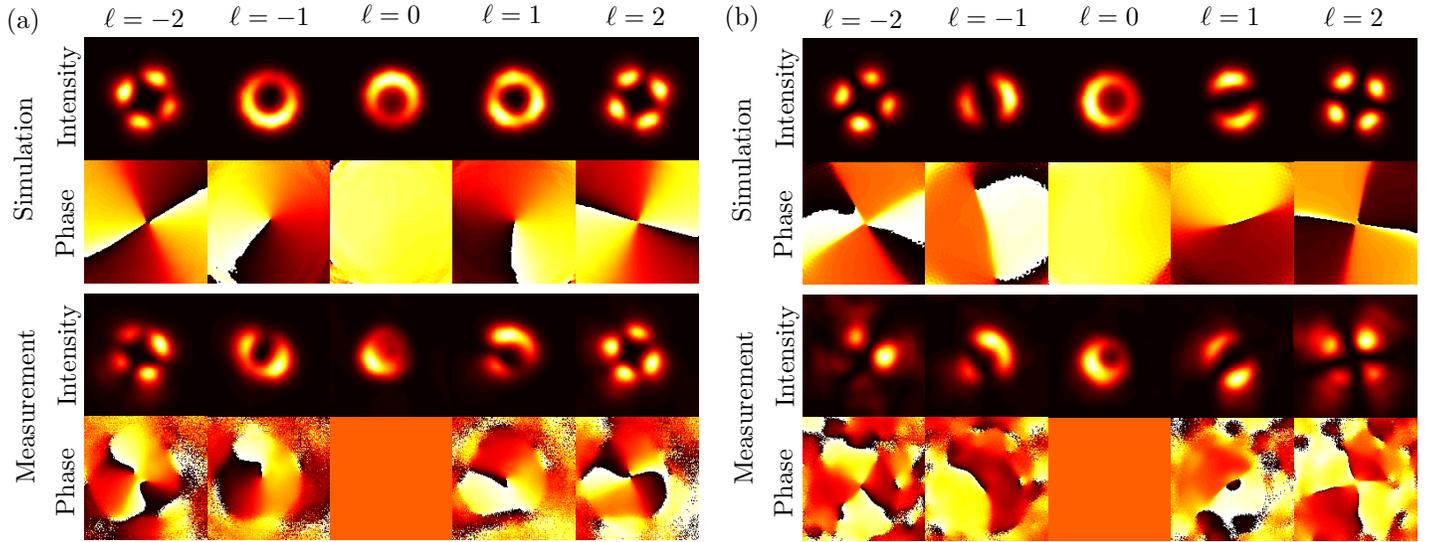
\caption{Comparison of output field of the BPM-simulations to the measured field from the printed structure for light coupled in different input waveguides at $\lambda=\SI{750}{\nano\metre}$. The intensity distribution is captured by a CMOS Camera while the phase distribution is measured relative to the $\ell=0$ mode by interference by coupling light in the two waveguides with different phase shifts. For a structure (a) with and (b) without an effective magnetic field}
\label{fig:output_field}
\end{figure}

For our initial comparison, a structure measuring \SI{4}{\milli\meter} in length was fabricated and single mode waveguides were utilized to couple light into the input side. The resulting spatial intensity and phase distribution on the output side was recorded. 
Upon direct comparison with the simulation in Figure~\ref{fig:output_field}a, numerous similarities are apparent. 
Notably, multiple modes are mixed in both the simulation and measurement.
This is most evident by the four spots resulting from the combination of the $\ell=-2$ and $\ell=+2$ modes.
Additionally, the phase distribution displays vortices both in simulation and in measurement.
Based on the simulations and measured intensity distribution, it is evident that only the $\ell=0$ mode has intensity in the center, leading us to assume that this mode has a predominantly smooth or constant phase front without any phase vortices. 
Based on this assumption, we use the $\ell=0$ mode to specifically measure the relative phase to other modes with respect to the $\ell=0$ mode.

For a structure without the effective magnetic field, the modes at the output facet of the structure are LP-modes. 
For these modes there are no clear votrecies, instead neighboring regions with nonzero intensity have a phase shift of about $\pi$ to each other. 
Again, for the case without an effective magnetic field, the simulation data qualitatively match the measured data, as can be seen in Figure~\ref{fig:output_field}b.

\begin{figure}
\def\svgwidth{\columnwidth}
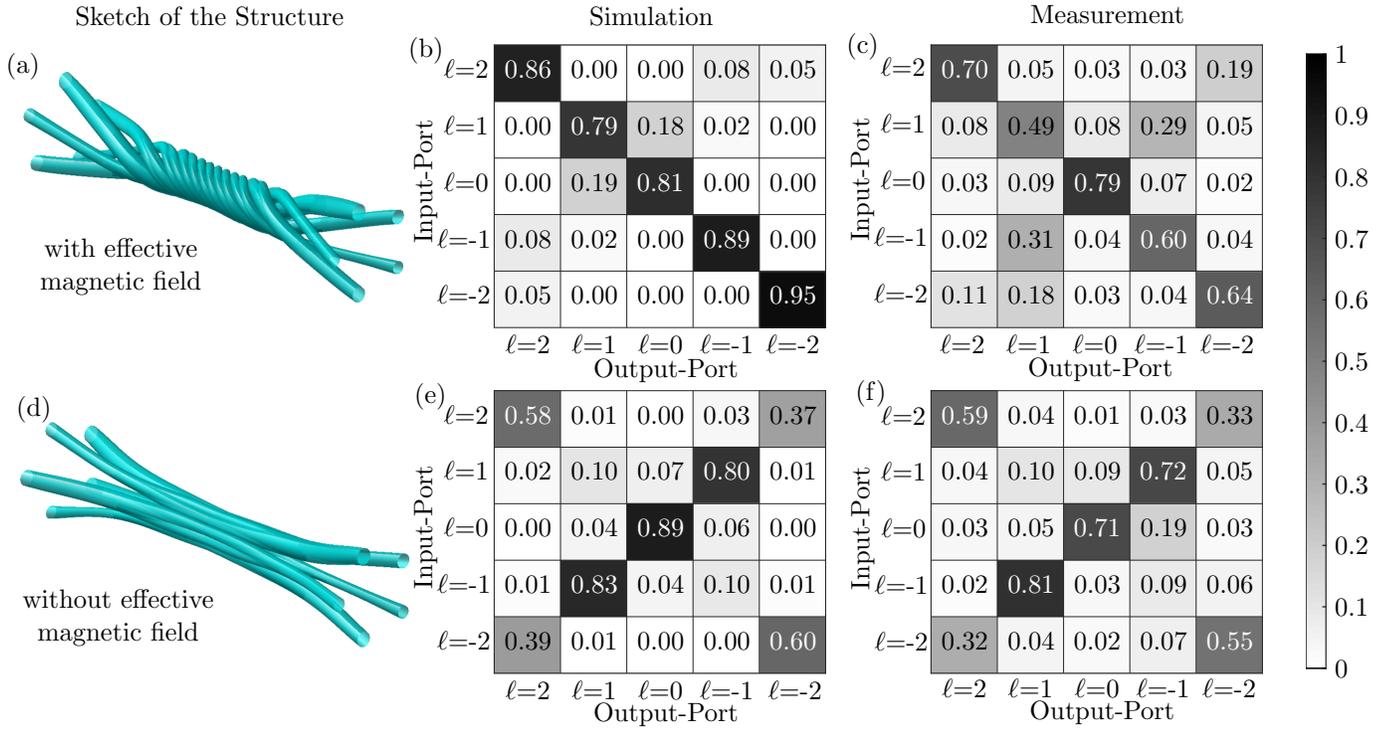
\caption{Characterisation of the mode conversion by a MUX/DEMUX setup. (a) and (d) Sketch of the structure. The demultiplexer is rotated against the multiplexer by $2\pi/5$ which causes crosstalk for LP-modes while OAM-modes are unaffected. (b) and (e) simulated and (c) and (f) measuremend crosstalk matrices of the structures for the case with and without an effective magnetic field at $\lambda=\SI{710}{\nano\metre}$, respectively.}
\label{fig:Mux_DeMux}
\end{figure}

To experimentally determine the mode selectivity of our device, a \SI{4}{\milli\metre} long multiplexing structure followed by a equivalent \SI{4}{\milli\metre} long demultiplexing structure is used (MUX/DEMUX)\cite{Mode-Selective_PL,PL_OAM_Mux}. 
In this way errors from misalignment of the OAM-mode can be excluded \cite{multiple_coherent_inputs}. 
To show that due to the effective magnetic field we are indeed multiplexing OAM-modes and not LP-modes the demultiplexer is rotated by $2\pi/5$ against the multiplexer. 
Since OAM modes are rotationally invariant except for phase, they can be accurately demultiplexed, while significant crosstalk is anticipated for LP-modes.
For this compression two structures following this MUX/DEMUX setup are considered one with and one without an effective magnetic field, as sketched in Figure~\ref{fig:Mux_DeMux}(a,d), respectively. 
The relative intensities at the output per input waveguide are plotted in  crosstalk matrices Figure~\ref{fig:Mux_DeMux}(b,c,e,f). 
The y-axis shows the input waveguide where the light was coupled into the multiplexer and the x-axis shows the relative intensities at the output waveguides after the demultiplexer. 
Due to this normalization, the sum of the intensities in each row is 1. 
In a perfect MUX/DEMUX setup, the crosstalk matrix would be the unity matrix.
Since the conversion to OAM-modes depends on the use of the effective magnetic field, Figure~\ref{fig:Mux_DeMux}(b,c) shows that the largest elements are on the diagonal of the crosstalk matrix. 
Without the effective magnetic field, only LP-modes are generated and demultiplexed, leading to high crosstalk entries outside the diagonal in Figure~\ref{fig:Mux_DeMux}(e,f). 
Again, simulation and measurement show qualitatively very similar results.

\section{Conclusions}
We have presented a waveguide structure that can be used to multiplex or demultiplex multiple OAM modes at different wavelengths simultaneously. 
By twisting the structure, the light experiences an effective magnetic field, so that the modes with equal absolute of OAM split energetically.
Because the structure is based on the adiabatic principle, it can be used intrinsically over a wide wavelength range, which we have demonstrated in simulations. 
The experiments on the fabricated structures are in good qualitative agreement with the simulations and thus demonstrate as proof of principle of the applicability for the generation or decomposition of OAM-modes.


\section{Materials and Methods}
\label{Materials_and_Methods}

The trajectories of the waveguides are parameterized in polar coordinates, where the radial component for all waveguides is given by $|\vec{r}|$. 
To split a waveguide into two waveguides in step 5, the angle $\alpha$ (see Figure~\ref{fig:modell_sketch}b) is slowly increased. 
For this purpose, instead of defining one waveguide, two equivalent waveguides are defined, which differ from the original one only in that $\pm\alpha$ is added to the polar coordinate, respectively. 
At the end facet, the 10 waveguides have a polar spacing of about $2\pi/10$ from each other. 
The area counted as the waveguide core is the area in which at least one of the parameterized waveguide trajectories lies. 
The area where multiple waveguides overlap is \underline{not} exposed multiple times.
The values to parameterize the steps we used for all simulations and the structure in the experiment are captured in Table~\ref{tab:StructureParameter}.

\begin{table}
\caption{parameter of the structure used in the simulations and experiment}
\label{tab:StructureParameter}
\begin{tabular}{c|l|c|c|c|c}
\hline\hline
& Description of the step 	& $s_\text{i}$			& $s_\text{f}$				& $\mu_s$	& $T_s$ \\ \hline
1   &   $\Delta	\dwg$ increase		& \SI{0}{\micro\metre}	& \SI{0.3204}{\micro\metre}	& 0.0541		& 0.0716	\\
2   &   $\abs{\vec{r}}$ decrease	& \SI{10}{\micro\metre}	& \SI{2.6}{\micro\metre}	& 0.1219		& 0.1604	\\
3   &   $\Omega$ increase		& \SI{0}{\milli\metre}$^{-1}$	& $2\pi$/(\SI{1748.7}{\milli\metre})	& 0.6119	& 0.2158	\\
4   &   $\Delta\dwg$ decrease		& \SI{0.3204}{\micro\metre}	& \SI{0}{\micro\metre}	& 0.6683	& 0.0983	\\
5   &   waveguides split up $\alpha$	& 0					& $2\pi/20$					& 0.6794		& 0.0736	\\
6   &   $\dwg$ decrease 			& \SI{2.6}{\micro\metre}	& \SI{1.7812}{\micro\metre}	& 0.8857	& 0.0858	\\
\hline\hline
\end{tabular}
\end{table}

The sample is fabricated in the negative photoresin IP-Dip (Nanoscribe) using a commercial direct laser writing (DLW) system (Nanoscribe Photonic Professional GT). 
The structure is written layer by layer, where each layer is stacked onto the other in the $z$-direction with a distance of \SI{250}{\nano\metre}. 
Each layer, is written line by line with a line distance of \SI{100}{\nano\metre}. 
The refractive index contrast of approximately $\Delta n\approx 0.008$ is achieved by choosing a high laser intensity for the area of the waveguide core (LaserPower 60\%) and low intensity for the surrounding area (LaserPower 24\%) close to the polymerization threshold (LaserPower 22\%) at a writing speed of \SI{20}{\milli\metre\per \second}. 
Thereby, a LaserPower of 100\% refers to a laser intensity of \SI{68}{\milli\watt} before the 63$\times$ focusing objective of the DLW system. 
For each layer the position and diameter of each waveguide is calculated according to the $z$-position by equation~(\ref{eq:FermiDirac}) with the parameters in Table~\ref{tab:StructureParameter}.
After the writing process, excess photoresist on the tip of the sample, that would cause distortions at the measurement, is removed by dipping the tip in PGMEA for a minute\cite{NNNSchulz}. 

For the measurement, laser light from a white light laser (NKT photonics) and a VARIA filter box is used. 
The light was linear polarized, expanded and send on a spatial light modulator (SLM). 
Afterwards, all light besides the first diffraction order of the blazed grating on the SLM is blocked. 
With a 20$\times$ objective (NA=0.4), the Fourier transformed hologram on the SLM is imaged on the input facet of the waveguide sample. 
With the hologram, we can choose the intensity and phase profile on the input facet. 
Intensity of the light from the output facet at the waveguide structure is imaged onto a CMOS camera.

Using the SLM, we couple into one of the $\ell=\{\pm 1,\pm 2\}$ waveguides and with different phase shifts in the $\ell=0$ waveguide simultaneously. 
Using the interference of the two modes, the relative phase for each pixel on the camera can be determined as the phase of a sinusoidal fit\cite{reconstruction_berry_curvature}.

\medskip

\medskip
\textbf{Acknowledgements} \par 
G.v.F. and J.S. acknowledge funding by the Deutsche Forschungsgemeinschaft through CRC/Transregio 185 OSCAR (project No.\ 277625399).

\section{Conflict of Interest}
The authors declare no conflict of interest.

\section{Data Availability Statement}
The data that support the findings of this study are available from the corresponding author upon reasonable request.

\medskip






\begin{figure}
\textbf{Table of Contents}\\
  \medskip
  \includegraphics[width=55mm]{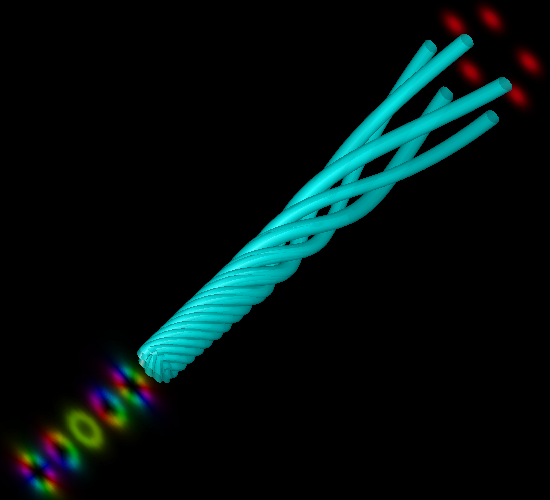}
  \medskip
  \includegraphics[width=110mm]{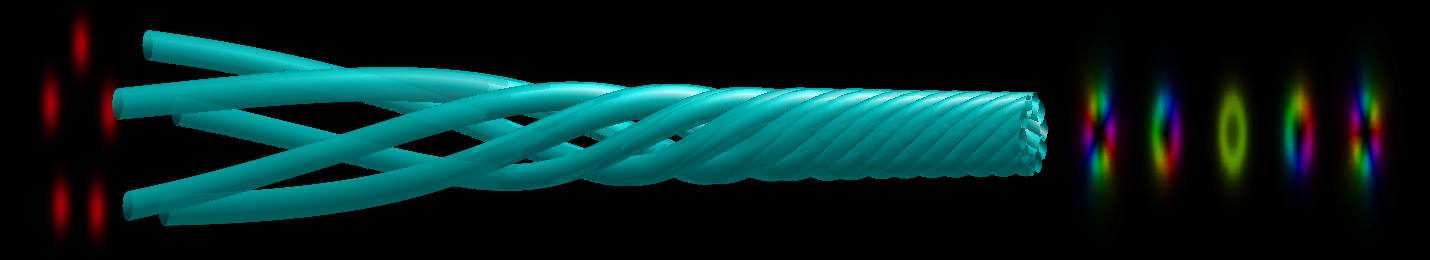}
  \medskip
  \caption*{A fiber based broadband OAM-(de)multiplexer is presented which is realized with a dielectric waveguide structure.
The structure design is based on the adiabatic principle, as utilized in photonic lanterns for highly efficient conversion of spatially separated single modes into eigenmodes.
An artificial magnetic field is introduced by twisting the structure to lift the degeneracy of the OAM modes.}
\end{figure}

\bibliographystyle{MSP}
\bibliography{Multiplexer}

\end{document}